\RequirePackage[2020-02-02]{latexrelease}
\documentclass[superscriptaddress,secnumarabic,
amssymb,amsmath,nobibnotes,aps,prd,showkeys,showpacs,nofootinbib]{revtex4}%
\usepackage{graphicx}
\usepackage{subfigure}
\usepackage{hyperref}
\usepackage{epsfig}
\usepackage{epsf}
\usepackage{bm}
\usepackage{amsmath}
\usepackage{amsfonts}
\usepackage{amssymb}
\usepackage{epstopdf}
\usepackage{natbib}%
\providecommand{\U}[1]{\protect\rule{.1in}{.1in}}
\newcommand{\be}{\begin{equation}}
\newcommand{\ee}{\end{equation}}
\usepackage{color}

\begin{document}

\title{Revisiting Barrow's Graduated Inflationary Universe: A Warm perspective}

\author{Subhra Bhattacharya}
\email{subhra.maths@presiuniv.ac.in}
\affiliation{Department of Mathematics, Presidency University, Kolkata-700073, India}

\begin{abstract}

It is presumed that thermal fluctuations present during inflationary epoch can make inflaton scalar field to interact with other fields resulting in the existence of a thermal component during the inflationary period. The presence of this thermal component assists structure formation and reduces reheating dependence as in the contemporary inflationary paradigm. This is known as warm inflation. In 1990 J . D. Barrow \cite{25} considered a scenario of inflation with matter field having a phenomenological equation of state of the type $p+\rho=\gamma\rho^{\lambda},~\gamma\neq 0$ and $\lambda$ constant. He called such inflationary scenarios as ``graduated inflation". In this work we reconsider the above equation of state in a scenario of warm inflation. Our aim would be to investigate and  understand whether such matter can also act as a viable candidate for warm inflation.

\end{abstract}

\maketitle
\section{Introduction}
\label{Intro}

The discovery of Cosmic Microwave Background Radiations (CMBR) by Penzias and Wilson in 1964-65 strongly suggested that the universe expanded from an early dense and hot state \cite{1,2}. Subsequently Alan Guth in 1981 \cite{3} put forward his theory of an exponentially expanding universe while in a supercooled vacuum state, which came to be known as the inflation. The inflationary paradigm is now the most compelling theory of the early universe. It not only solves some of the persistent cosmological problems, like the horizon, flatness and monopole problem, but also provides scale invariant nearly Gaussian density fluctuations as observed by CMBR. In the classical model of inflation the scalar field $\phi,$ also called the inflaton field, would slowly roll down along a flat potential. At the end of this inflation the inflaton field would oscillate about a minimum potential creating the friction necessary for transfer of energy from the inflaton field to other fields. This was the followed by nucleosynthesis and primordial structure formation \cite{5}. (For a review on inflation refer \cite{4}). 

Although the usual inflationary model does not assume the effects of energy dissipation, yet it has been speculated that dissipative effects, resulting in interactive energy exchanges between the various existing matter fields might have been in action during the process of inflation \cite{6, 7}. This new scenario where one considered interactions between the inflaton and radiation through a dissipative term was called the ``warm inflation" in contrast to the classical ``cold inflation". Warm inflation is ``warm" due to the continuous dissipation of energy from inflaton to the radiation field, which creates a thermal component during the process of inflation. Due to the continuous heat generation, reheating was no longer essential at the end of inflation, hence providing a smoother mechanism for primordial large scale structure formation \cite{7,8,9}. Further, the presence of the thermal bath could explain the background thermal fluctuations and their differential density fluctuation \cite{6,8,9,10,11,hall1,hall2}. The interaction induced thermal fluctuations is mathematically expressed using a dissipative term $\Gamma.$ The dissipative term influenced the inflationary period by elongating it, where slow roll conditions could be relaxed \cite{12}. 

Over the past two decades variety of work has been done on the warm inflation. Warm inflation has been been successfully modelled with quadratic, hilltop, quartic and hybrid potentials \cite{13}. They have been studied in canonical and non-canonical scalar field models \cite{14} and in the context of generalized Chaplygin gas models \cite{15}. Over the past few years, several warm inflation models have been discussed in the literature in a variety of diverse theoretical scenarios, like the multifield models \cite{16}, on the swampland criteria \cite{17}, in the Dirac-Born-Infeld non-canonical scalar fields \cite{18} etc.

The motivation of this article is to study warm inflation scenario in the presence of inflaton field generated due to matter with a phenomenological equation of state (that has been sometimes called inhomogeneous equation of state, \cite{no1}). In the perspective of recent observational cosmology, it is predicted that we live in a universe that is undergoing accelerated expansion \cite{22}. Mathematically such a universe requires the existence of matter with negative pressure, known as the Dark Energy (DE). The DE represents about 70\% of the total energy in the universe \cite{23}. Physicists have been perplexed with DE since its formulation. Several different theories from gravity sector modification to introduction of interacting dynamics between the several matter components have been explored, (for a complete review refer \cite{24}). One such probable explanation for DE was provided by the introduction of phenomenological equation of state for DE. Such phenomenological e.o.s can be expressed by a general relation given by: $p+\rho=A\rho^{\alpha},$ where $p$ is the pressure density and $\rho$ the energy density of matter with $A$ and $\alpha$ being constants. Such inhomogeneous e.o.s was previously used by Barrow in \cite{25} to extend the behaviour of power law inflationary universes. It was called the ``graduated inflationary universe". He showed that it was possible to represent a wide class of inflationary universes by using the above $\rho-p$ relation. Later, similar equation of states were used to predict the thermodynamics of DE universes \cite{26} and in the context of emergent universes \cite{27}.  They have also been extensively used to explain the emergence of phantom epoch \cite{pm} and nature of future singularities \cite{28,no1,no2}. More recently they have been used to describe a finite time strong singularity free universe \cite{29}. In this work we shall probe the above equation of state and try to find how they can influence the warm iflationary dynamics. We shall use the e.o.s as inflaton energy density and effectively develop the ensuing warm inflationary dynamics in the presence of a constant dissipative term. 

The paper is organised as follows: Section 2 deal with a general discussion on the basic idea of warm inflation with its salient features and guiding mathematical relations. In section 3 we shall describe our model, providing analytical solutions for the relevant physical and dynamical terms. Section 4 shall use numerical evaluations for the analytical results and correspondingly provide graphical representation of the relevant parameter behaviour. Finally the paper ends with a brief conclusion and discussion in section 5.

\section{Warm Inflation: A brief recapitulation }

Considering a flat FRW universe, with an inflaton field $\phi$ having potential $V(\phi),$ and a radiation field with energy density $\rho_{r},$  the Friedmann's equations give,
\begin{align}
3H^{2}=\kappa(\rho_{\phi}+\rho_{r})\label{fd1}\\
-2\dot{H}=\kappa(\rho_{\phi}+p_{\phi}+\frac{4}{3}\rho_{r})\label{fd2}
\end{align}
where $\rho_{\phi}=\frac{1}{2}\dot{\phi}^{2}+V(\phi)$ and $p_{\phi}=\frac{1}{2}\dot{\phi}^{2}-V(\phi)$ are the inflaton energy density and pressure. Constant $\kappa=8\pi G=\frac{8\pi}{m_{p}^{2}}$ with $G$ the Gravitational constant and $m_{p}$ the Plank's mass. $H=\frac{\dot{a}}{a}$ is the Hubble parameter. The dynamical conservation equations corresponding to the two energy densities are given as:
\begin{align}
\dot{\rho_{\phi}}+3H(\rho_{\phi}+p_{\phi})=-\Gamma\dot{\phi^{2}}\label{cv1}\\
\dot{\rho_{r}}+4H\rho_{r}=\Gamma\dot{\phi^{2}}.\label{cv2}
\end{align}
Here $\Gamma$ is the dissipative coefficient, such that $\Gamma>0$ and dot represents differentiation with respect to time $t$. $\Gamma$ can be constant or it can be a function of temperature $T$ or field $\phi$ or a function of both $\phi$ and $T,$ that is, $\Gamma(\phi,T)$ \cite{6,8,9,hall1,hall2}. Equation (\ref{cv1}) can be rewritten as the inflaton conservation equation from $\rho_{\phi}=\frac{1}{2}\dot{\phi}^{2}+V(\phi)$ as:
\begin{equation}
\ddot{\phi}+(3H+\Gamma)\dot{\phi}+V_{\phi}=0.\label{cv3}
\end{equation} 
with $V_{\phi}=\frac{\partial V}{\partial\phi}.$ 

The equations (\ref{fd1}-\ref{cv3}) are simplified by considering the ensuing process to be i) slow rolling in inflaton field and ii) quasi-stable in radiation field. The slow roll conditions $\dot{\phi}^{2}\ll V(\phi)$ and $\ddot{\phi}\ll(3H+\Gamma)\dot{\phi},$ imply scalar field dominates the potential during the inflationary era. Hence, from equation (\ref{cv3}), one obtains $\dot{\phi}=-\frac{V_{\phi}}{(3H+\Gamma)}.$ The second condition imply that radiation field although co-exists with the inflaton field, will be dominated by it and that the production of radiation during this epoch will be quasi-stable. This essentially mean $\rho_{\phi}\gg\rho_{r}$ and $\dot{\rho_{r}}\ll\rho_{r},~\dot{\rho_{r}}\ll\Gamma\dot{\phi}^{2}.$ Thus from (\ref{fd1}) and (\ref{cv3}) one obtains $3H^{2}=\kappa \rho_{\phi},$ and $\rho_{r}=\frac{\Gamma\dot{\phi}^{2}}{4H}$ \cite{9,hall1, hall2,10}.

Using the $\rho_{r}\simeq C_{r}T^{4}=\frac{\Gamma\dot{\phi}^{2}}{4H}$ the temperature of the thermal bath $T$ can be obtained as:
\begin{equation}
T=\left(\frac{3R\dot{\phi}^{2}}{4 C_{r}}\right)^{\frac{1}{4}}.\label{T}
\end{equation}
Here constant $C_{r}=\frac{g_{*}\pi^{2}}{30}$ is the Stephen-Boltzman constant with $g_{*}$ being the number of relativistic degrees of freedom corresponding to the radiation field \cite{6,7,8} and $R=\frac{\Gamma}{3H}$ is redefined as the dissipation rate with $R\gg 1$ corresponding to strong dissipative regime (that is dissipation rate is higher than the expansion rate) and $R\ll 1$ the weak dissipative regime ($\Gamma\ll 3H$).

Using the above equations the dimensionless slow roll markers $\epsilon=-\frac{\dot{H}}{H^{2}}$ and $\eta=-\frac{\ddot{H}}{H\dot{H}}$ can be obtained as $\epsilon_{w}=(1+R)\epsilon;~\eta_{w}= (1+R)\eta.$ These apart there are two other parameters $\beta,$ that signify the $\phi$ dependence of the dissipation and $\iota$ corresponding to the temperature dependence of the potential \cite{10,11}. Therefore the relevant set of slow-roll markers for warm inflation are now four in number and given by:
\begin{equation}
\epsilon=\frac{1}{2\kappa}\left(\frac{V_{\phi}}{V}\right)^{2};~\eta=\frac{1}{\kappa}\left(\frac{V_{\phi\phi}}{V}\right);~\beta=\frac{1}{\kappa}\left(\frac{V_{\phi}\Gamma_{\phi}}{\Gamma V}\right);~\iota=\frac{TV_{\phi T}}{V_{\phi}}.\label{para}
\end{equation}
The validity of the slow-roll corresponds to all the above parameters smaller than $(1+R).$ Subscripts represent corresponding variable differentiation.

Since the density fluctuations in warm inflation are thermal in nature, it is predicted that both entropy and scalar perturbations would be significant in warm inflation. However it was eventually established in \cite{hall2} that only scalar perturbations remain dominant at large scales. Thus it is only relevant to study the power spectra due to the scalar density fluctuations, which now being thermal in nature is given by $\mathcal{P_{S}}=\frac{H^{3}T}{\dot{\phi}^{2}}\sqrt{1+R}$ \cite{hall1,hall2}, with the scalar spectral index $n_{s}-1=\frac{d\ln\mathcal{P_{S}}}{d\ln k}$ given by \cite{moss2}
\begin{equation}
n_{s}-1=\frac{\dot{\mathcal{P_{S}}}}{H\mathcal{P_{S}}}\simeq-\frac{9R+17}{4(1+R)^{2}}\epsilon-\frac{9R+1}{4(1+R)^{2}}\beta+\frac{3}{2(1+R)}\eta.\label{ns}
\end{equation}
Here $k$ is the co-moving wave number and $k=aH$ is the time when dissipation damps fluctuations, which being taken same as the horizon cross over time \cite{moss2}. The tensor scalar modes are same as in case of classical inflation and is given by $\mathcal{P_{T}}=8H^{2}$ \cite{moss1, moss2} with spectral index $n_{t}-1=-\frac{2\epsilon}{1+R}.$ The tensor to scalar ratio $r=\frac{\mathcal{P_{T}}}{\mathcal{P{S}}}$ is given by the relation
\begin{equation}
r\simeq\frac{16\epsilon H}{T(1+R)^{\frac{5}{2}}}.\label{r}
\end{equation}

\section{Warm inflation and a general in-homogeneous equation of state}  

We consider the general inhomogeneous equation of state for the inflaton energy density as given by \cite{25}
\begin{equation}
p_{\phi}+\rho_{\phi}=A\rho_{\phi}^{\alpha}\label{inh},
\end{equation} with $\alpha$ and $A$ constants such that $\alpha\neq 1,$  (in (\ref{inh}) replacing $\alpha=1$ gives us the usual perfect fluid equation of state). In \cite{25} Barrow showed that this simple $\rho-p$ relation could generate a wide class of inflationary models just by the variation of some constant parameters. 

We use (\ref{inh}) in the conservation equation for inflaton (\ref{cv1}) together with the relation $\dot{\phi}^{2}=\rho_{\phi}+p_{\phi},$ and obtain 
\begin{equation}
\rho_{\phi}=\gamma^{\frac{1}{1-\alpha}}\left[\ln\left(\frac{a}{a_{i}}\right)\right]^{\frac{1}{1-\alpha}}\label{rho}
\end{equation}
for $\alpha\neq 1,$ with $R$ being dimensionless constant and $a$ the scale factor and $a_{i}$ some constant factor. Here $\gamma=3A(1+R)(\alpha-1)$ is a constant. 
For $\rho_{\phi}>0$ we would require $\gamma>0,$ which can be attained under the following restrictions:
\begin{itemize}
\item $\alpha>1,~A>0.$ In this case at $a=a_{i},~\rho_{\phi}\rightarrow\infty.$ 
\item $\alpha<1,~A<0.$ Here we have for $a=a_{i},~\rho_{\phi}\rightarrow 0.$
\item $\frac{1}{1-\alpha}=${\it even integer} irrespective of the sign of $A.$ Which confines $\alpha$ in the interval $(\frac{1}{2},1)\cup(1,\frac{3}{2}).$ Evidently for $\alpha<1,$ at $a=a_{i},~\rho_{\phi}\rightarrow 0$ while for $\alpha>1,~\rho_{\phi}\rightarrow\infty.$
\end{itemize}
Here we may note that the $\frac{p_{\phi}}{\rho_{\phi}}=-1+A\rho_{\phi}^{\alpha-1}.$  Given, that for $A>0$ we have constrained $\alpha>1$ which gives the vacuum DE equation of state for $\rho_{\phi}\rightarrow 0$. While for $A<0$ and $\alpha <1$ we get a highly phantom type fluid with diverging equation of state for $\rho_{\phi}\rightarrow 0.$ 

For the special case of $\alpha =1$ we can obtain $\rho_{\phi}=\rho_{0}a^{-3A(1+R)},$ with $\rho_{0}$ being some integration constant. At the beginning of inflation when radiation fields are weak, and can be neglected, we use $3H^{2}=\rho_{\phi}$ together with the equation (\ref{cv1}) to obtain $\rho_{\phi}= \frac{4C_{0}}{3A^{2}(1+R)^{2}t^{2}}$ where $C_{0}$ is the constant of integration. From the above two expressions for $\rho_{\phi}$ it is evident that for $\alpha=1$ with negligible radiation density the scale factor $a$ will be given by $a=a_{i}t^{\frac{2}{\gamma_{0}}}$ where $\gamma_{0}=3A(1+R)$ is a constant dependent on $R$ and $A.$ Evidently this is the usual power law inflation. 

Similarly for any general $\alpha,$ after neglecting the radiation density we obtain:
\begin{itemize}
\item[{\bf C1:}] For the choice of the parameter $\alpha=\frac{1}{2},~A<0$ we obtain:
\begin{enumerate}
\item[{\bf R1:}] $\rho_{\phi}=\rho_{0}exp\left[\frac{\gamma t}{\sqrt{3}}\right]$
\item[{\bf A1:}] $a(t)=a_{i}exp\left[\frac{\rho_{0}^{1/2}}{\gamma}exp\left[\frac{\gamma t}{\sqrt{3}}\right]\right]$
\end{enumerate}

\item[{\bf C2:}] For $\alpha<\frac{1}{2},~A<0$ and $\alpha>1,~A>0$ we obtain the solution as: 
\begin{itemize}
\item[{\bf R2:}] $\rho_{\phi}=\rho_{0}(\beta \gamma t)^{\frac{2}{1-2\alpha}}$
\item[{\bf A2:}] $a(t)=a_{i}exp\left[\rho_{0}^{1/2}\gamma^{\frac{1}{1-2\alpha}}\left(\beta t\right)^{\frac{2(1-\alpha)}{1-2\alpha}}\right]$ with $\beta= \frac{2\alpha-1}{2\sqrt{3}(\alpha-1)}.$
\end{itemize}
\item[{\bf C3:}] And for $\frac{1}{2}<\alpha<1,~A<0$
\begin{itemize}
\item[{\bf R3:}] $\rho_{\phi}=\rho_{0}\left[\frac{(1-2\alpha)^2}{12(\alpha-1)^2}\gamma^{2}t^{2}\right]^{\frac{1}{1-2\alpha}}$
\item[{\bf A3:}] $a(t)=a_{i}exp\left[-\frac{\rho_{0}^{1/2}}{\sqrt{3}}\left(\frac{2\alpha-1}{2(1-\alpha)}\right)\left(\frac{(1-2\alpha)^{2}}{12(\alpha-1)^{2}}\gamma^{2}\right)^{\frac{1}{2(1-2\alpha)}}t^{\frac{2(1-\alpha)}{1-2\alpha}}\right].$
\end{itemize}
\end{itemize} 
As in \cite{25}, we see that by considering (\ref{inh}) in the presence of an interaction with radiation energy density we could obtain several categories of solution parametrized by the exponent $\alpha.$ From the case C2 we have  $a(t)\rightarrow\infty$ as $t\rightarrow\infty$ while from C3 we see that at $t=0,~a=0$ and proceeds to $a\rightarrow a_{i}$ at $t\rightarrow\infty.$ 

Using the relation $\dot{\phi}=\frac{d\phi}{da}aH$ and $H$ approximated with $\rho_{\phi}$ we obtain the scalar field $\phi$ in terms of the scale factor $a$ as follows:
\begin{equation}
\left(\frac{\phi-\phi_{0}}{2}\right)^{2}=\frac{3 A}{\gamma}\ln\left(\frac{a}{a_{i}}\right)\label{sf}
\end{equation}
which gives the potential as:
 \begin{equation}
V(\phi)=\frac{\gamma^{\frac{2}{1-\alpha}}}{(3A)^{\frac{1}{1-\alpha}}}\left(\frac{\phi-\phi_{0}}{2}\right)^{\frac{2}{1-\alpha}}-\frac{A}{2}\frac{\gamma^{\frac{2\alpha}{1-\alpha}}}{(3A)^{\frac{\alpha}{1-\alpha}}}\left(\frac{\phi-\phi_{0}}{2}\right)^{\frac{2\alpha}{1-\alpha}}\label{pot}.
\end{equation}
The above relations completely describe the initial inflaton field. Here $V(\phi)\propto \phi^{s}$ where $s$ is a fractional number. Thus we see that at the beginning, the considered energy density gives an intermediate inflation.

\subsection{Radiation energy density and the inflaton field}

As inflation proceeds radiation energy density gets significant and can be obtained from (\ref{cv2}) and $p_{\phi}+\rho_{\phi}=\dot{\phi}^{2}$ with $\dot{\rho_{r}} <<\rho_{r}$ as: 
\begin{equation}
\rho_{r}=\frac{3}{4}RA\gamma^{\frac{\alpha}{1-\alpha}}\left[\ln\left(\frac{a}{a_{i}}\right)\right]^{\frac{\alpha}{1-\alpha}}.\label{rhor}
\end{equation}
Now using the results for $\rho_{\phi}$ and $\rho_{r},$ in the equation (\ref{fd1}) and (\ref{fd2}) we get:
\begin{align}
3H^{2}&=\gamma^{\frac{1}{1-\alpha}}\left[\ln\left(\frac{a}{a_{i}}\right)\right]^{\frac{1}{1-\alpha}}+\frac{3}{4}RA\gamma^{\frac{\alpha}{1-\alpha}}\left[\ln\left(\frac{a}{a_{i}}\right)\right]^{\frac{\alpha}{1-\alpha}}\label{h}\\
-2\dot{H}&=(1+R)A\gamma^{\frac{\alpha}{1-\alpha}}\left[\ln\left(\frac{a}{a_{i}}\right)\right]^{\frac{\alpha}{1-\alpha}}.\label{hdt}
\end{align}

As the field evolves the initial scalar field due to inflaton energy density is now replaced with the total energy density of the inflaton field and the radiation field. So that we have $\dot{\phi^{2}}=(1+R)A\gamma^{\frac{\alpha}{1-\alpha}}\left[\ln\left(\frac{a}{a_{i}}\right)\right]^{\frac{\alpha}{1-\alpha}}.$ This gives the new scalar field potential as: 
\begin{equation}
V(\phi)=V_{0}\left[(\phi-\phi_{0})^{2}-\frac{2}{3(\alpha-1)^{2}}\right]\left[(\phi-\phi_{0})^2-\frac{R}{(1+R)(\alpha-1)^{2}}\right]^{\frac{\alpha}{1-\alpha}}\label{potr}
\end{equation}   
where $V_{0}=\left(\frac{\gamma(\alpha-1)}{4}\right)^{\frac{1}{1-\alpha}}$ and $\phi-\phi_{0}=\frac{2\sqrt{3A(1+R)}}{\gamma}\sqrt{\gamma\ln\left(\frac{a}{a_{i}}\right)+\frac{3}{4}RA}.$ This also gives an intermediate inflation. Comparing equations (\ref{pot}) and (\ref{potr}) we see that in the presence of radiation energy density the scalar field potential varies by some factor proportional to the dissipation rate $R$ and the exponent $\alpha$ of the the inflaton energy density. We see that due to the presence of the dissipation factor, potential field will roll down slower as compared to the that in the absence of dissipation. This is consistent with warm inflationary dynamics.

Using the above equations one can also consider the dynamics of the above warm inflationary scenario. The relevant slow roll parameters are obtained as:
\begin{equation}
\epsilon=\frac{\frac{3}{2}(1+R)A}{\frac{3}{4}RA+\gamma\ln\left(\frac{a}{a_{i}}\right)}\qquad \eta=\frac{3(1+R)A}{\frac{3}{4}RA\alpha+\gamma\ln\left(\frac{a}{a_{i}}\right)}\qquad \beta=\frac{\frac{3}{2}(1+R)A}{\frac{3}{4}RA+\gamma\ln\left(\frac{a}{a_{i}}\right)}\label{sr}
\end{equation}
Using the above slow roll parameters and the equations (\ref{ns}) and (\ref{r}) we obtain the scalar spectral index $n_{s}$ and the tensor to scalar ratio $r$ as:
\begin{align}
n_{s}&=1-\frac{27A}{4\gamma \ln\left(\frac{a}{a_{i}}\right)+3AR}+\frac{18A\alpha}{4\gamma \ln\left(\frac{a}{a_{i}}\right)+3AR\alpha}\\
r&=16\left(\frac{4g_{*}\pi^{2}A^{3}(\gamma \ln\left(\frac{a}{a_{i}}\right))^{\frac{\alpha}{1-\alpha}}}{90R}\right)^{\frac{1}{4}}\left(\frac{3}{(1+R)^{3}\left(4\gamma \ln\left(\frac{a}{a_{i}}\right)+3RA\right)}\right)^{\frac{1}{2}}.\label{nsr}
\end{align}

\subsection{Numerical and Graphical Results}

It may be noted that the term $\ln\left(\frac{a}{a_{i}}\right)$ gives us an accurate estimate of the number of e-folds $N$ until the end of inflation. Since the results obtained in the above analysis are all functions of $\ln\left(\frac{a}{a_{i}}\right)$ they can, in general be represented as a function of the e-folding number $N$ where $N$ can be obtained in terms of time $t$ using  the results  A1 - A3 for the cases C1 - C3. This gives an estimate on the range of parameters that can give 60 e-folds of inflation before it ends. 

In Figure 1 we have plotted the values of $N(t)=\ln\left(\frac{a}{a_{i}}\right)$ corresponding to the equations A1-A3 and for various values of the dissipative factor $R$. Figure 1(a) gives the e-folding number  with respect to the time $t$ corresponding to equation A2. This figure shows that for $\alpha>1$ and $A>0$, a weak dissipative factor $R<<1$ takes less time to get to the 60 e-folds mark as compared to a stronger dissipation, i.e. $R>1.$ In figure 1(b) we have plotted the e-folding number corresponding to equation A1 for $\alpha=0.5.$ Here the scale factor $a(t)$ is dependent on a double exponential term, as a result inflation proceeds faster as compared to in scenario A2. Further a higher dissipative regime will give a faster inflation than a weaker dissipative regime as is evident in the figure. Figure 1(c) represents the case $\alpha<0.5$ and $A<0$ for equation A2. Here the scale factor $a(t)$ evolves as an exponential of some positive power of $t$ that is $a\propto e^{t^{\delta}}$ with $\delta>0$ as result inflation proceeds fast with faster rate for higher dissipative scenario. Figure 1(d) corresponds to equation A3. The scale factor $a\propto e^{-t^{-\delta}},$ as a result $a\rightarrow a_{i}$ as $t\rightarrow\infty.$ 

\begin{figure}
\centering
\subfigure[]{\includegraphics[width=0.46\textwidth]{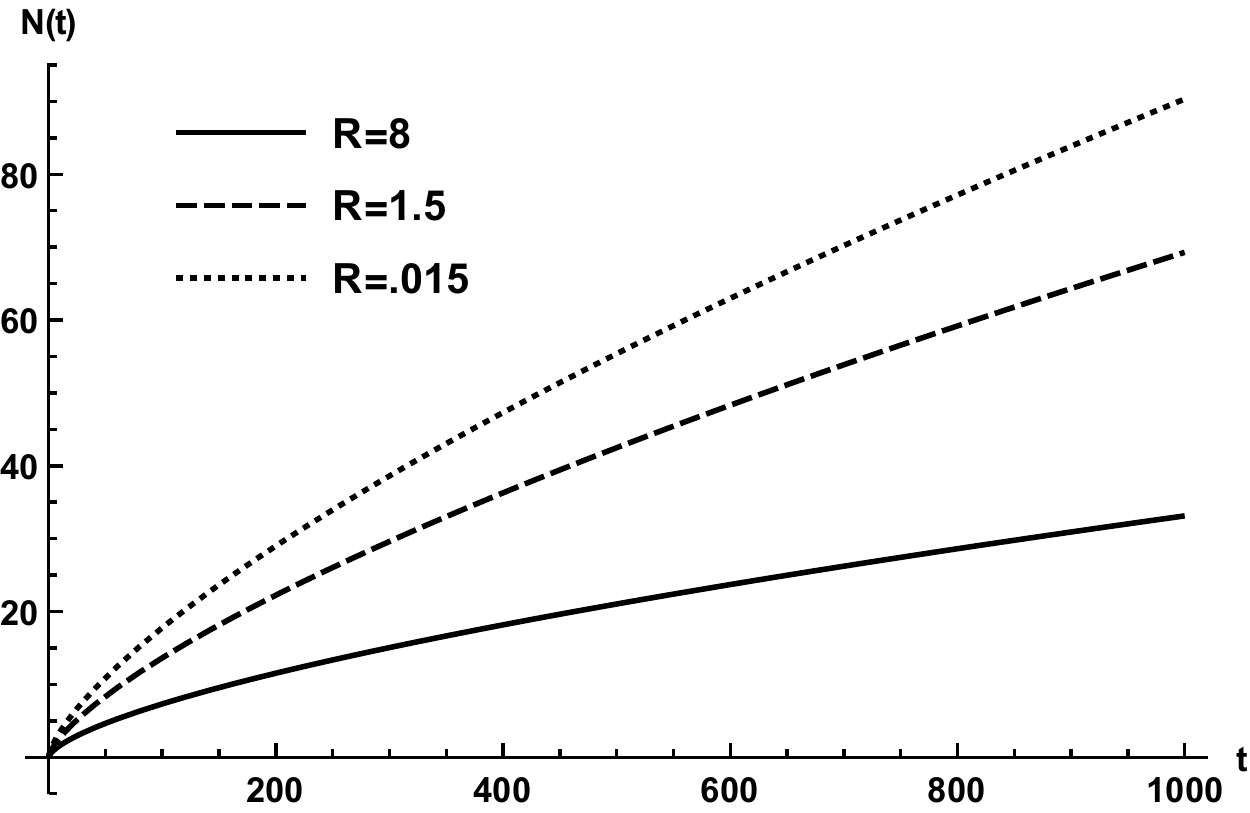}}
\subfigure[]{\includegraphics[width=0.46\textwidth]{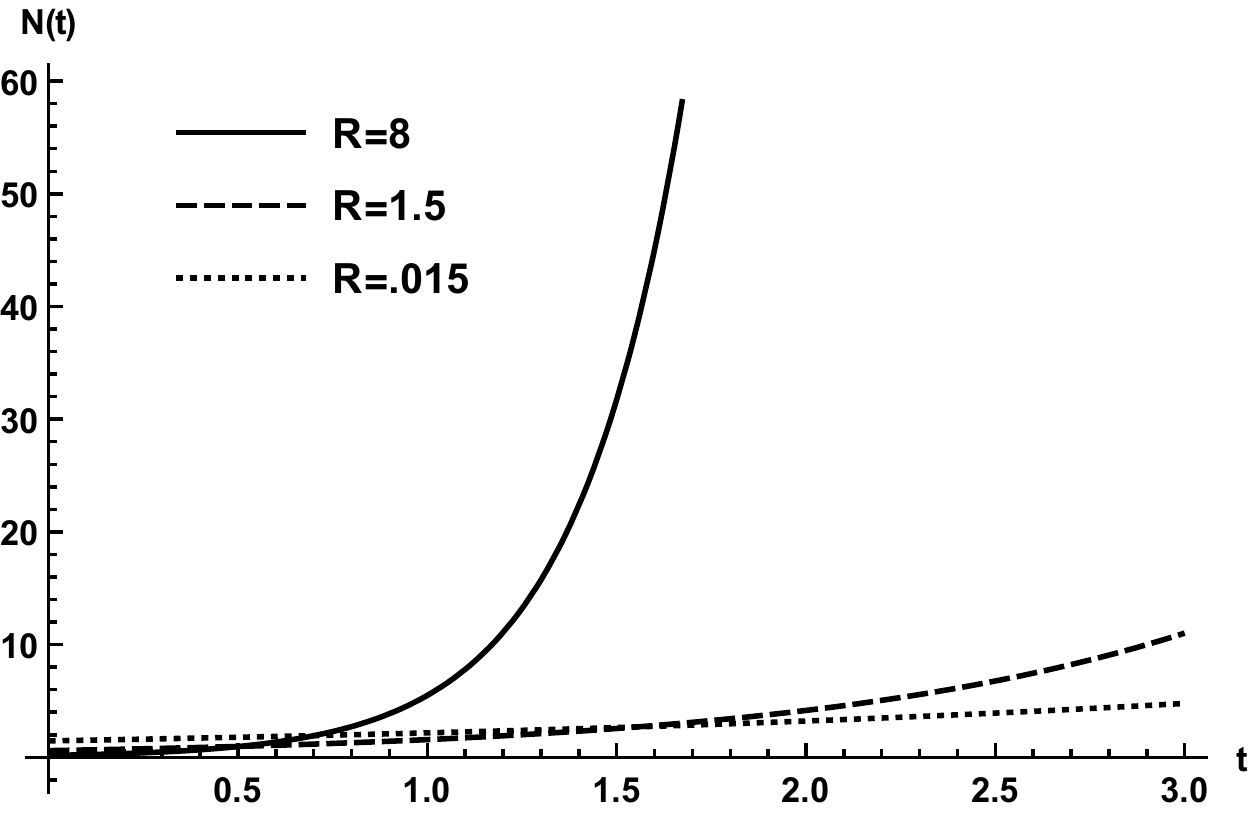}}
\subfigure[]{\includegraphics[width=0.46\textwidth]{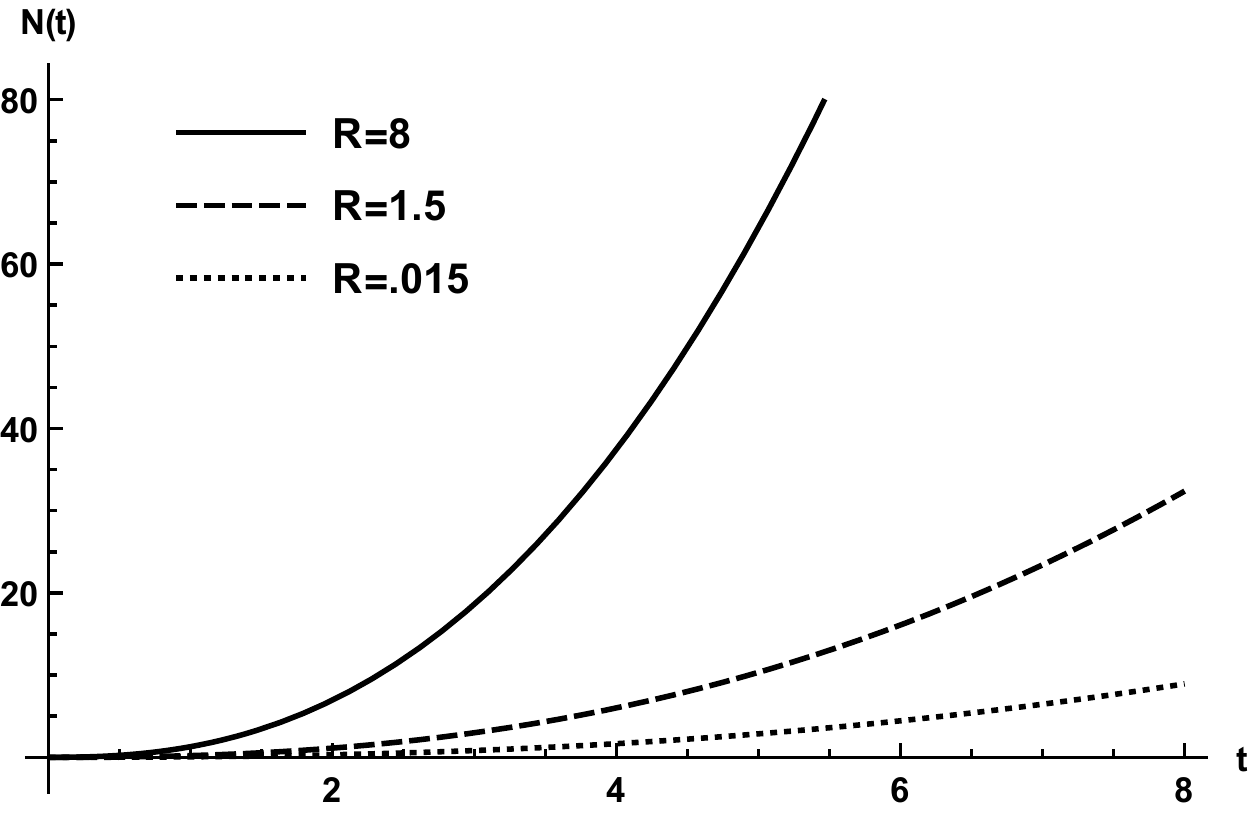}}
\subfigure[]{\includegraphics[width=0.46\textwidth]{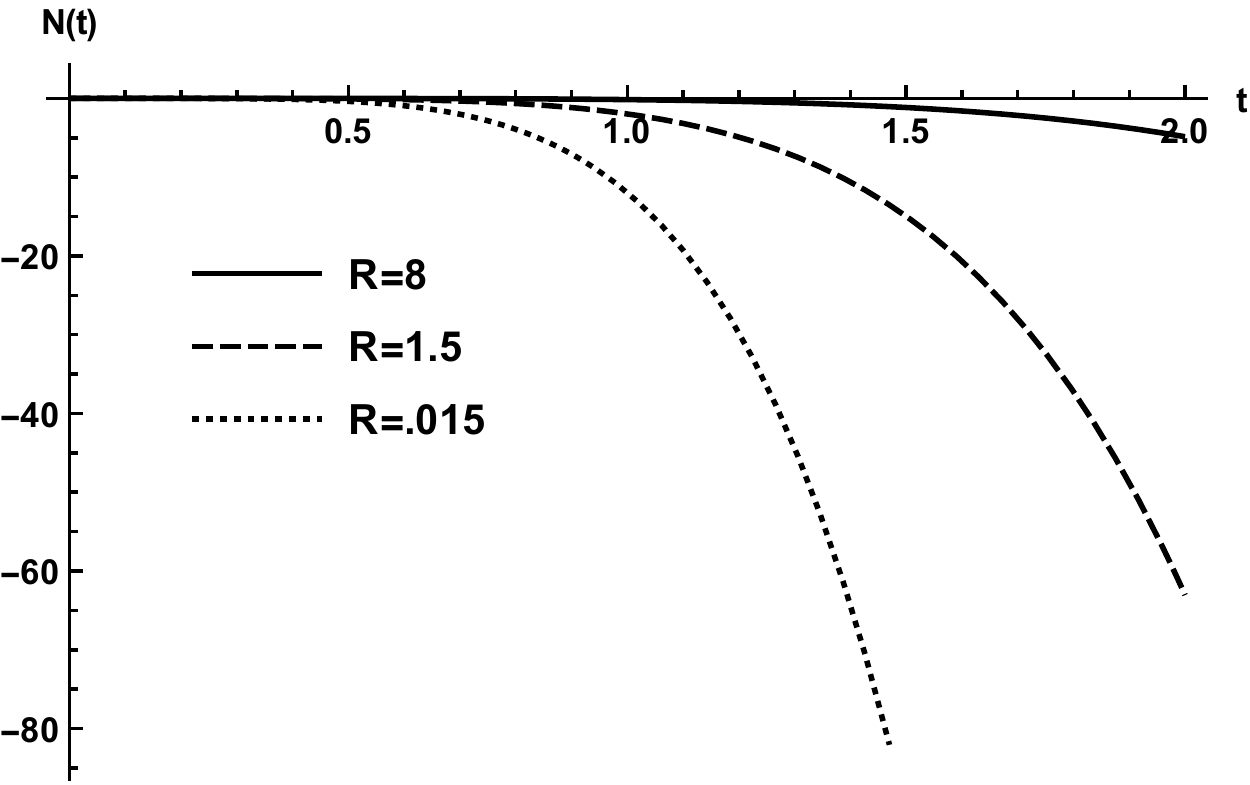}}

\caption{Evolution of $N(t)=\ln\left(\frac{a}{a_{i}}\right)$ corresponding to (a)  Equation {\bf A2} for $\alpha=1.95>1$ and $A=0.6$ (b) Equation {\bf A1} for $\alpha=0.5$ and $A=-0.6$ (c) Equation {\bf A2} for $\alpha=0.15<\frac{1}{2}$ and $A=-0.6$ (d) Equation {\bf A3} for $\frac{1}{2}<\alpha=0.75<1$ and $A=-0.6$}
\end{figure}

\begin{figure}
\centering
\subfigure[]{\includegraphics[width=0.46\textwidth]{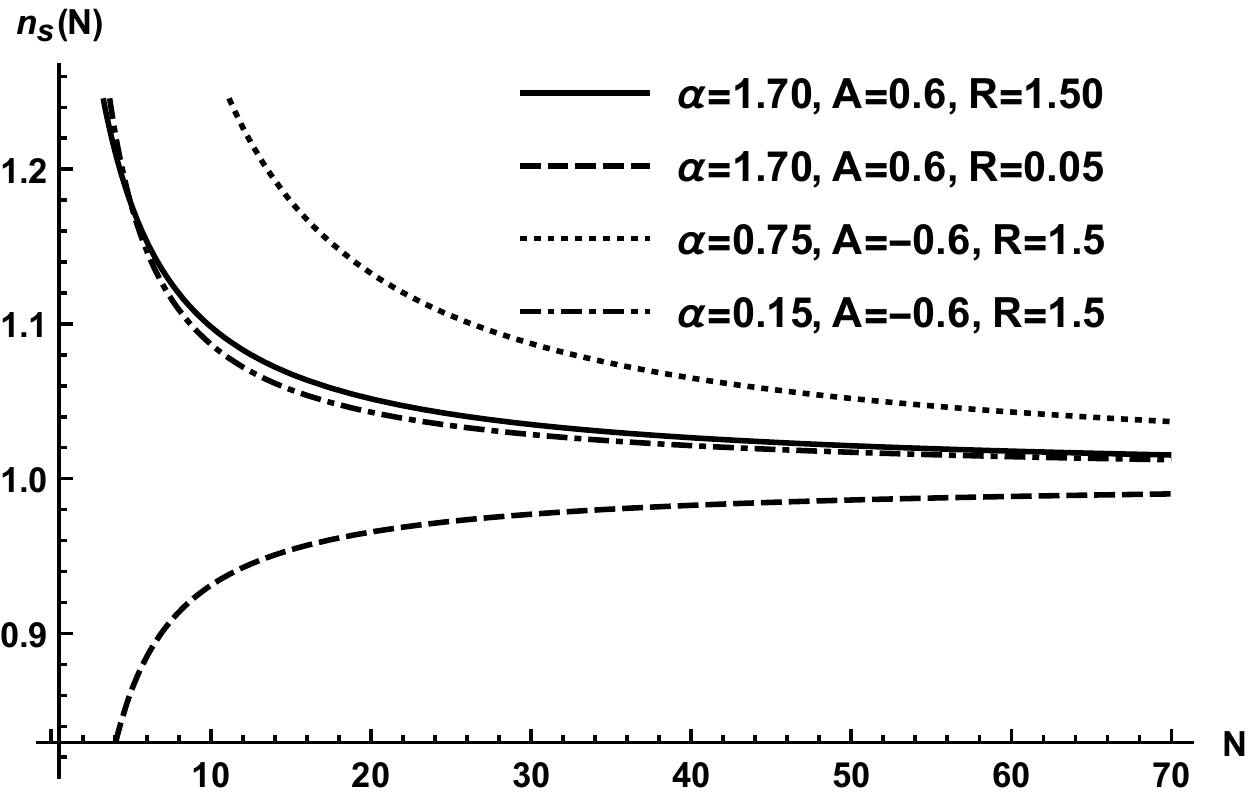}}
\subfigure[]{\includegraphics[width=0.46\textwidth]{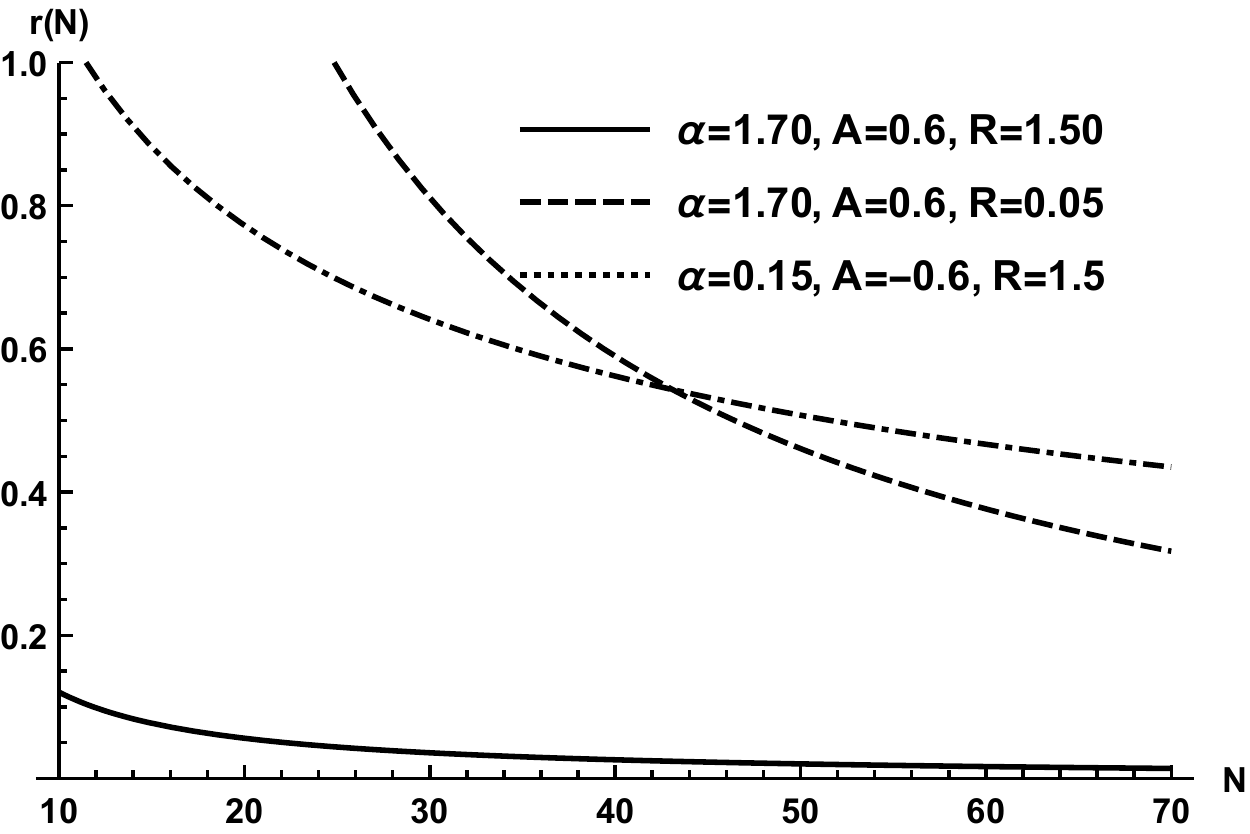}}
\caption{(a) $n_{s}$ plotted as a function of $N(t)$  (b) $r$ plotted as a function of $N(t)$}
\end{figure}

In figure 2 the scalar spectral index $n_{s}$ is plotted for the four cases considered in A1-A3. Recent Planck data constrains the scalar spectral index $n_{s}$ for classical inflation models to $n_{s}\simeq 0.9659\pm 0.0040$ while the bounds on the tensor to scalar ratio $r$ is put at $r < 0.072.$ Figures 2a and 2b shows the behaviour of $n_{s}$ and $r$ corresponding to the specific parameter choices. The parameters have been chosen so that they represent the cases C1 - C3 and for $R>1$ and $R<1.$ The graphical representations of $n_{s}$ and $r$ show that the model performs best for the parameter ranges of $\alpha>1$ and $A>0.$

Figures 3 show the values of the scalar field $\phi$ and the potential $V(\phi)$ against the e-folding number $N$. From both figures we see that the values of $\phi$ to be increasing while $V(\phi)$ decreasing as is expected from a slowly rolling inflaton field. 
\begin{figure}
\centering
\subfigure[]{\includegraphics[width=0.46\textwidth]{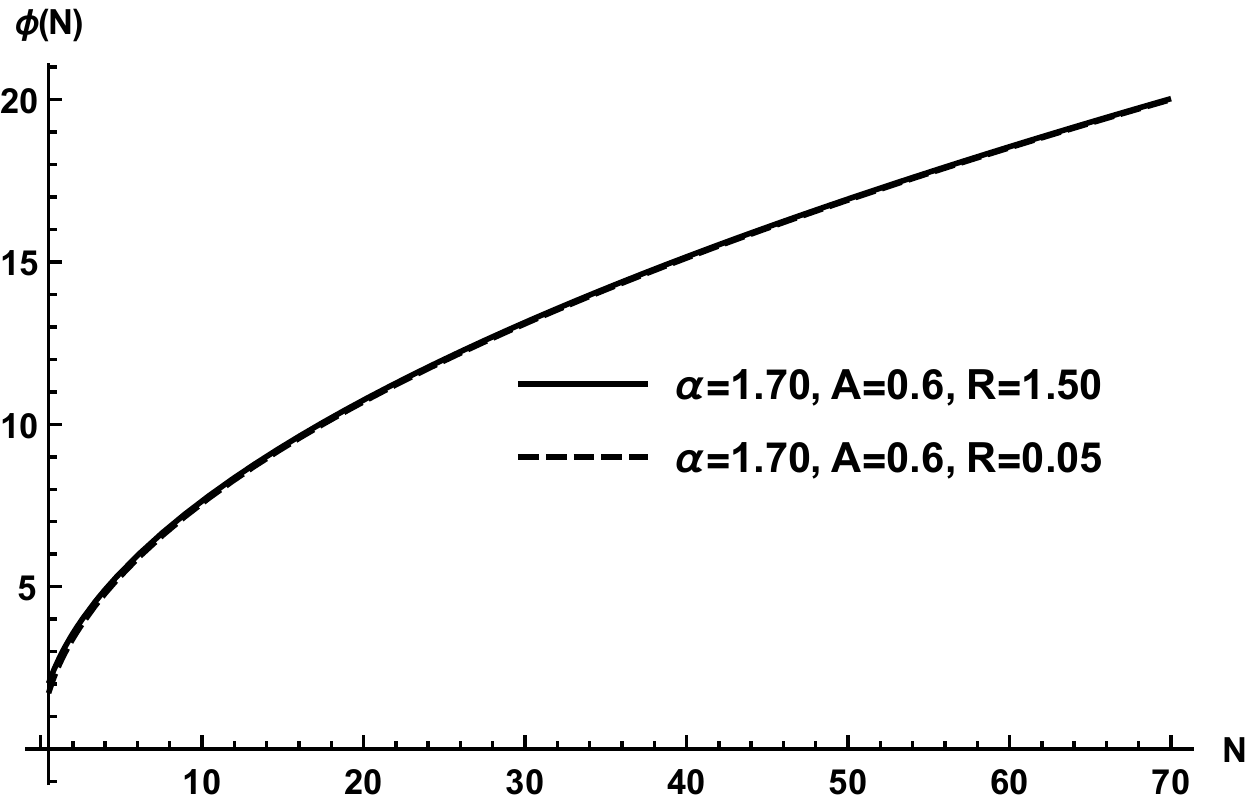}}
\subfigure[]{\includegraphics[width=0.46\textwidth]{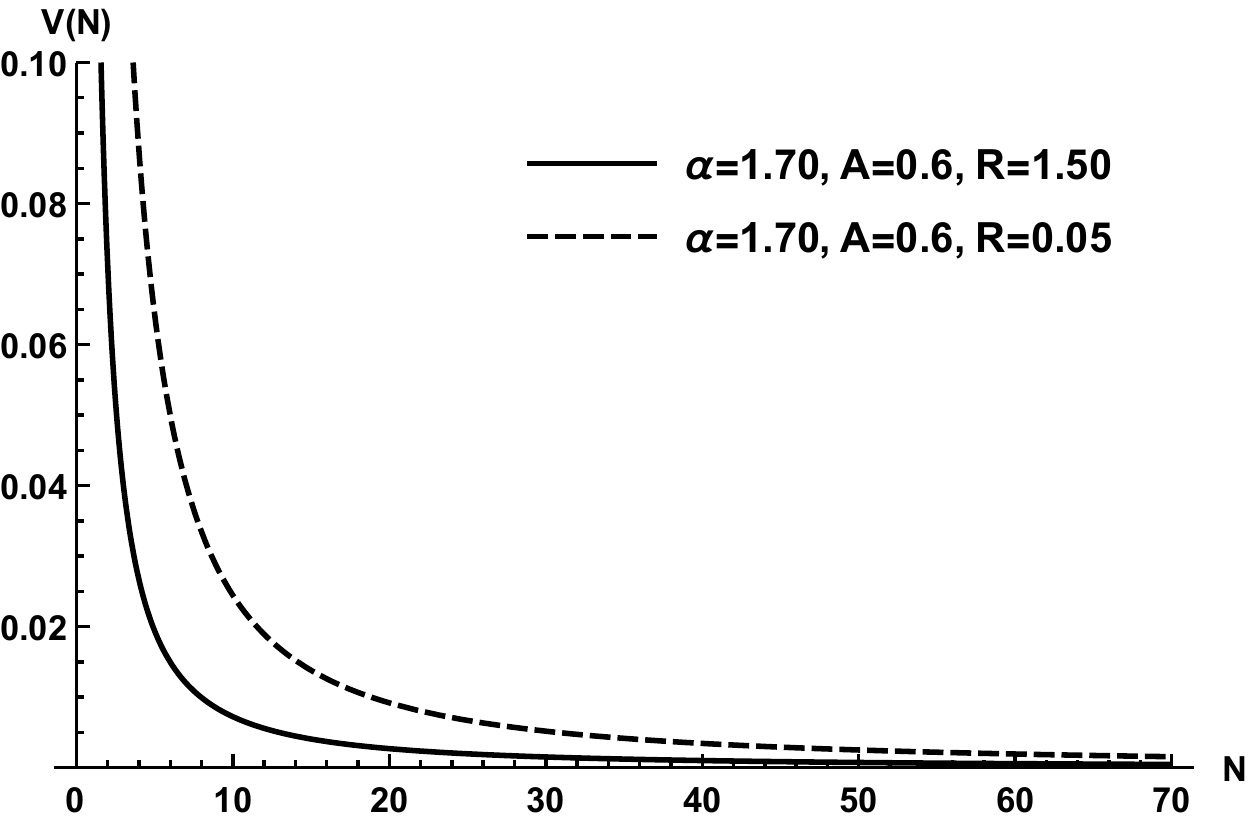}}
\caption{(a)$V(\phi)$ plotted as a function of $N(t)$ (b) $\phi$ plotted as a function of $N(t)$}
\end{figure}

In figure 4a we have shown the general behaviour of $V(\phi)$ against $\phi$ for a specific choice of parameter. It may be stated that similar plots can be obtained for other parameter choices. In fig 4b, we have plotted $n_{s}$ against $r$ for a particular case of $\alpha>1$ and $A>0.$ It may be noted that$n_{s}$ and $r$ are not always consistent to the Planck 2018 prescribed bounds. Consistent results are obtained in the case of $\alpha>1$ and $A>0$ corresponding to both $R>1$ and $R<1.$

\begin{figure}
\centering
\subfigure[]{\includegraphics[width=0.46\textwidth]{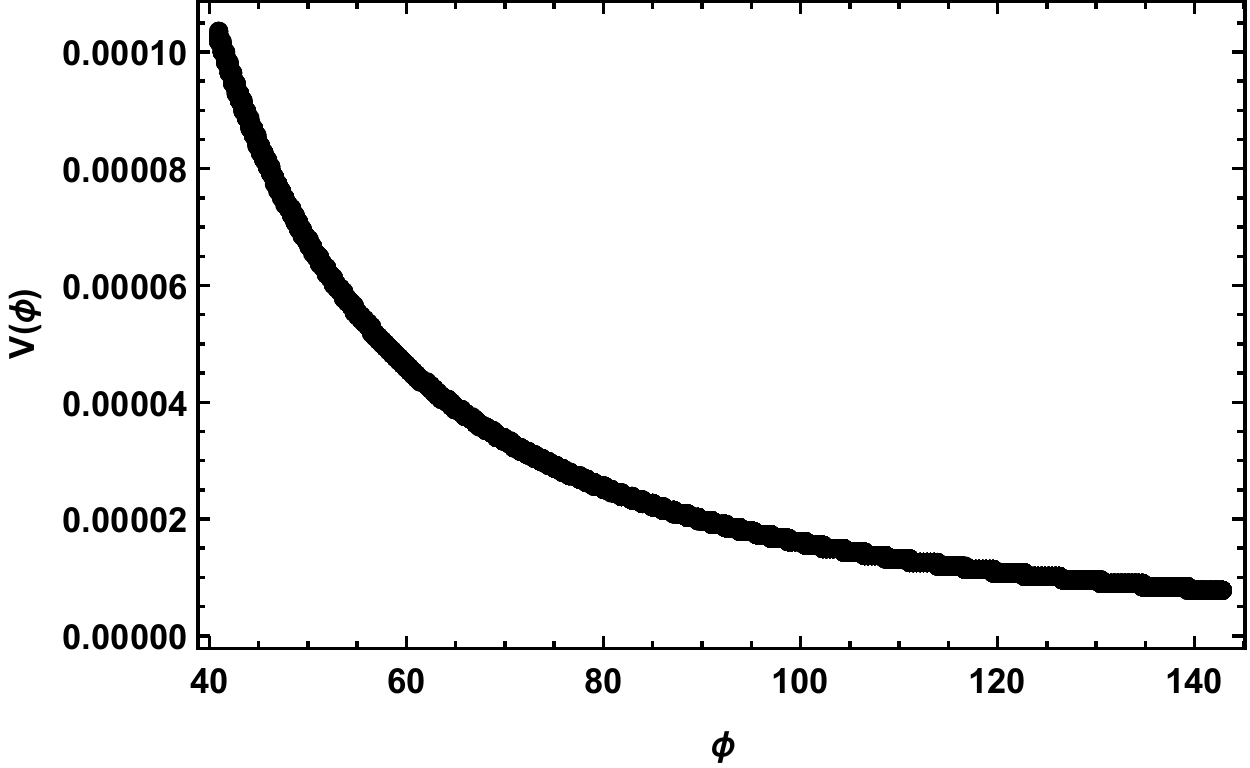}}
\subfigure[]{\includegraphics[width=0.46\textwidth]{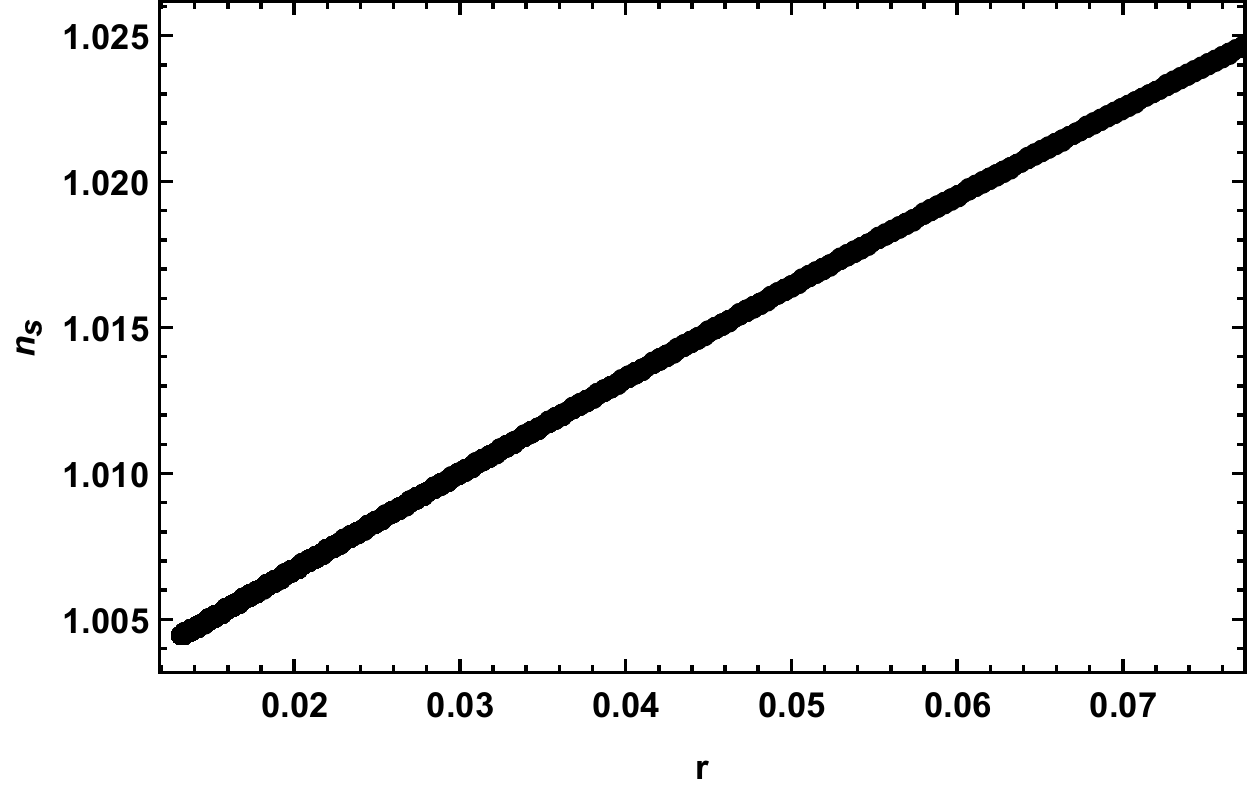}}
\caption{(a) $V(\phi)$ vs $\phi$ plotted for $\alpha=1.95,~A=0.6,~R=2.5$ (b) $n_{s}$ vs $r$ for $\alpha=1.95,~A=0.6,~R=2.5$}
\end{figure}

\section{Discussions}

In this article we have considered a scenario of warm inflation where the inflaton field couples with radiation energy via a non-trial coupling term, called the dissipative constant. The inflaton energy density driving the warm inflation was previously discussed in \cite{25} and was found to exhibit a wide class of inflationary universes. In the presence of coupling with radiation field we find that the corresponding field exhibits intermediate inflation where the scale factor varies as the exponential of time $t^{f}$ where $f>0$ or $f<0.$ Further by special choice of the parameters we could also obtain a scenario power law inflation and a double exponential inflation. In all the cases we have evaluated the relation between the scalar field and the field potential. Graphical representations of the models exhibit behaviour consistent to inflation. Further we have shown how the presence of the dissipative term affects the field potential making the field to roll down slower. The behaviour of the $n_{s}$ vs. $r$ plot is also consistent with inflationary models. Thus using a non-trivial coupling between a general phenomenological equation of state inflaton field and radiation energy density we could successfully obtain a model of warm intermediate inflation.

\section{Acknowledgments}
 SB acknowledges Dr. Tanwi Bandyopadhyay for her suggestions and Prof. S. Chakraborty for his encouragements. SB thanks IUCAA for their research facility where this work was initiated.


\end{document}